\documentclass[aps,prl,twocolumn,groupedaddress,showpacs]{revtex4}

\usepackage{graphicx}
\usepackage{dcolumn}
\usepackage{bm}
\def\Vec#1{\mbox{\boldmath $#1$}}
\def\Vecsf#1{\mbox{\boldmath ${\mathsf #1}$}}

\begin{document}


\title{Presence of 3$d$ Quadrupole Moment in LaTiO$_3$ 
Studied by $^{47,49}$Ti NMR}


\author{Takashi Kiyama}
\affiliation{Department of Physics, Graduate School of Science, Nagoya University, 
Nagoya 464-8602, Japan}
\author{Masayuki Itoh}
\affiliation{Department of Physics, Graduate School of Science, Nagoya University, 
Nagoya 464-8602, Japan}


\date{\today}

\begin{abstract}
$^{47,49}$Ti NMR spectra of LaTiO$_3$ are reexamined and 
the orbital state of this compound is discussed. 
The NMR spectra of LaTiO$_3$ taken at 1.5 K under zero external field 
indicate a large nuclear quadrupole splitting. 
This splitting is ascribed to the presence of the rather large quadrupole moment 
of 3$d$ electrons at Ti sites, 
suggesting that the orbital liquid model proposed for LaTiO$_3$ 
is inappropriate. 
The NMR spectra are well explained by 
the orbital ordering model expressed approximately as 
$1/\sqrt{3} (d_{xy}+d_{yz}+d_{zx})$ 
originating from a crystal field effect. 
It is also shown that most of the orbital moment is quenched. 
\end{abstract}

\pacs{75.30.-m, 75.50.Ee, 76.60.-k}

\maketitle


The orbital degree of freedom has become 
an important topic in the physics of 
strongly correlated electron systems\cite{Tokura-Nagaosa}. 
One of the recent topics is 
a quantum effect in the orbital ordering 
dominated by electron correlations. 
Especially, Mott insulators with $t_{2g}$ 
degenerated orbitals are prospective in this viewpoint, 
since the Jahn-Teller (JT) coupling for $t_{2g}$ orbitals is weaker 
than $e_g$ states and the large degeneracy enhances 
the quantum effect. 

The RTiO$_3$ (R:rare earths) system is fascinating 
from the above viewpoint\cite{Keimer,KhaliullinMaekawa,Kikoin,
KhaliullinOkamoto,Ulrich}. 
RTiO$_3$ has an orthorhombically distorted perovskite structure 
(space group: $P_\mathrm{bnm}$), in which the electronic configuration of 
a Ti$^{3+}$ ion is ${t_{2g}}^1$. 
While several studies have reported that 
YTiO$_3$ takes an orbital ordering state from 
both experimental and theoretical aspects\cite{Ulrich,
KhaliullinOkamoto,Mizokawa,Sawada,Akimitsu,Itoh1,Nakao}, 
the orbital state in LaTiO$_3$ is 
now under controversy\cite{KhaliullinMaekawa,
Keimer,Mochizuki2,Cwik,Hemberger}. 
LaTiO$_3$ undergoes 
$G$-type antiferromagnetic ordering below $T_\mathrm{N} \sim$ 150 K 
with the reduced ordered moment of 0.46 $\mu_B$\cite{Meijer}. 
The question is what interaction lifts the degeneracy of 
threefold $t_{2g}$ orbitals in the ground state. 
The octahedra of TiO$_6$ in LaTiO$_3$ 
exhibit small distortions. 
One would expect, at first sight, quadruply degenerated single-ion ground states 
represented by fictitious angular momentum $\tilde{j}$= 3/2 
due to spin-orbit (SO) interaction, 
which have unquenched orbital moment. 
This scenario is consistent with the observed reduced magnetic moment. 
However, Keimer \textit{et al.} suggested that the SO interaction 
is not dominantly working in this system 
from the isotropic spin wave dispersion\cite{Keimer}. 
Under these circumstances, a picture of strongly fluctuating orbital states was 
proposed\cite{Keimer,KhaliullinMaekawa}, 
followed by a refined calculation 
on the orbital excitations\cite{Kikoin}. 
On the other hand, Mochizuki and Imada 
have successfully explained the physical properties of LaTiO$_3$ 
by an orbital state 
expressed approximately as $1/\sqrt{3} (d_{xy}+d_{yz}+d_{zx})$\cite{Mochizuki2}. 
Subsequently, corresponding distortions of the TiO$_6$ octahedra 
have been reported from detailed structural data\cite{Cwik,Hemberger}.

In this paper, we report the results of analyses on NMR measurements for LaTiO$_3$. 
The NMR spectra show that the quadrupole moment of a 3$d$ electron in 
LaTiO$_3$ is rather large and unfavorable for the orbital liquid model, while 
the spectra are well explained by the $1/\sqrt{3} (d_{xy}+d_{yz}+d_{zx})$-like 
orbital ordering state with appropriate parameters. 

\begin{figure}[b]
\begin{center}\leavevmode
\includegraphics[width=0.77\linewidth]{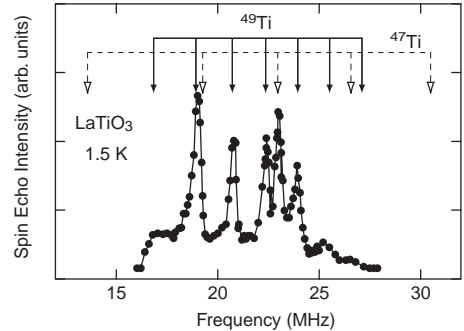}
\caption{ $^{47,49}$Ti NMR spectra under zero external field at 1.5 K 
in the antiferromagnetic state of LaTiO$_3$\cite{Itoh1}. 
Solid circles represent experimental data, and 
solid and dashed arrows represent the calculated resonance positions 
for $^{49}$Ti and $^{47}$Ti nuclei, respectively. 
}\label{figure1}
\end{center}
\end{figure}

Experimental procedures are mentioned in Ref.10. 
It is known that the magnetic properties are particularly 
sensitive to the oxygen content $\delta$ in LaTiO$_{3+\delta}$\cite{Meijer}. 
We used powdered samples with the N\`eel temperature 
$T_\mathrm{N} \sim$ 147 K ($\delta \sim$ 0.0) for NMR measurements. 
Frequency-swept NMR spectra of LaTiO$_3$ 
were taken point by point of frequency 
by using a super-heterodyne coherent pulsed spectrometer at 1.5 K 
in the antiferromagnetically ordered state 
under zero external field. 
As for the properties of $^{47}$Ti and $^{49}$Ti nuclei, 
the nuclear spin $^{47}I =\frac{5}{2}$ and $^{49}I =\frac{7}{2}$, 
the quadrupole moment $^{47}Q$ = 0.29$\times$10$^{-24}$ and 
$^{49}Q$ = 0.24$\times$10$^{-24}$ cm$^2$, 
and the gyromagnetic ratio $^{47}\gamma_N/2\pi$ = 2.4000 
and $^{49}\gamma_N/2\pi$ = 2.4005 MHz/T are used in this paper\cite{MetallicShifts}. 

Figure 1 shows the frequency-swept $^{47,49}$Ti NMR spectra 
in the antiferromagnetically ordered state of LaTiO$_3$, 
which were previously reported in Ref.10. 
These spectra are consistent with the data in Ref.17. 
Each peak of the spectra is assigned 
for $^{47}$Ti and $^{49}$Ti nuclei as shown by the arrows in Fig. 1. 
The attribution of the peaks 
to the $^{47}$Ti and $^{49}$Ti nuclei is 
the same as previous reports\cite{Itoh1,Furukawa}, 
although the parameters such as the quadrupole frequency $\nu_Q$ and 
the asymmetry parameter of the electric field gradient (EFG) $\eta$ 
are not unique to reproduce the spectra. 

The NMR spectrum of Ti nuclei is determined 
by the hyperfine interaction. 
The hyperfine interaction $\mathcal{H}$ in the 3$d^1$ transition metal ion is 
expressed as\cite{Abragam} 
\begin{eqnarray}
&\lefteqn{}& \!\!\!\! \mathcal{H} = \mathcal{H}_\mathrm{mag}+\mathcal{H}_\mathrm{el} 
 = - \gamma_N \hbar {\Vec H}^\mathrm{eff} \cdot {\Vec I} + 
\frac{eQ{\Vec I} \cdot \Vecsf{V}^\mathrm{eff} \cdot {\Vec I}}{2I(2I-1)} ,  \\
&\lefteqn{}& \!\!\!\! \mathcal{H}_\mathrm{mag} = 
2\mu_B \gamma_N \hbar \langle r^{-3} \rangle_\mathrm{mag} 
[- \kappa \Vec{S}\cdot\Vec{I} + \Vec{L}\cdot\Vec{I} + \frac{2}{21} \nonumber \\
& & \!\!\! \times \{L(L+1)\Vec{S}\cdot\Vec{I} - 
\frac{3}{2}(\Vec{L}\cdot\Vec{I})(\Vec{L}\cdot\Vec{S})
-\frac{3}{2}(\Vec{L}\cdot\Vec{S})(\Vec{L}\cdot\Vec{I})\}] \nonumber \\
& & \!\!\!\! \,\, = 2\mu_B \gamma_N \hbar \langle r^{-3} \rangle_\mathrm{mag} 
[- \kappa \Vec{S}\cdot\Vec{I} + \Vec{L}\cdot\Vec{I} 
- \frac{2}{21} \Vec{S} \cdot \Vecsf{q} \cdot \Vec{I}],  \\
&\lefteqn{}& \!\!\!\! \mathcal{H}_\mathrm{el} = 
\mathcal{H}_\mathrm{el}^\mathrm{on} 
+ \mathcal{H}_\mathrm{el}^\mathrm{out}, \\
&\lefteqn{}& \!\!\!\! \mathcal{H}_\mathrm{el}^\mathrm{on} 
= \frac{e^2 Q}{21I(2I-1)}\langle r^{-3} \rangle_\mathrm{el}
\{3(\Vec{L}\cdot\Vec{I})^2+\frac{3}{2}(\Vec{L}\cdot\Vec{I}) \nonumber \\
& & \!\!\!\! -L(L+1)I(I+1)\} 
= \frac{e^2 Q}{21I(2I-1)}\langle r^{-3} \rangle_\mathrm{el}
[\Vec{I} \cdot \Vecsf{q} \cdot \Vec{I}],  \\ 
&\lefteqn{}& \!\!\!\! \mathcal{H}_\mathrm{el}^\mathrm{out} 
= \frac{(1-\gamma_{\infty})eQ}{6I(2I-1)} 
\sum_{\alpha\beta} V_{\alpha\beta}^\mathrm{out} [\frac{3}{2}(I_{\alpha}I_{\beta} +
I_{\beta}I_{\alpha})-\delta_{\alpha\beta}I^2] \nonumber \\
& & \!\!\!\! \quad =(1-\gamma_{\infty})\frac{eQ}{2I(2I-1)} 
[\Vec{I} \cdot \Vecsf{V}^\mathrm{out} \cdot \Vec{I}], 
\end{eqnarray} 
where 
$\mathcal{H}_\mathrm{mag}$ represents the magnetic hyperfine interaction, 
$\mathcal{H}_\mathrm{el}$ the electric hyperfine interaction, 
$\mathcal{H}_\mathrm{el}^\mathrm{on}$ the electric hyperfine interaction with 
on-site $d$ electrons, 
$\mathcal{H}_\mathrm{el}^\mathrm{out}$ the electric hyperfine interaction with 
outside ions, 
${\Vec H}^\mathrm{eff}$ the effective internal magnetic field, 
$\Vecsf{V}^\mathrm{eff}$ the effective electric field gradient (EFG) tensor, 
$\hbar$ the Planck's constant, 
$\mu_B$ the Bohr magneton, 
$\langle r^{-3}\rangle_\mathrm{mag}$ and $\langle r^{-3}\rangle_\mathrm{el}$ 
the expectation values of $r^{-3}$ 
for the Ti $3d$ electron, 
$\kappa$ a parameter for the Fermi contact interaction 
due to a core-polarization effect, 
$\gamma_\infty$ the Sternheimer antishielding factor, 
{\boldmath $S$} the spin, {\boldmath $L$} 
the orbital momentum, {\boldmath $I$} 
the nuclear spin operator, 
$\Vecsf{q}$ the electron quadrupole moment tensor 
of which the components are defined as 
$q_{\alpha \beta}\equiv 
\frac{3}{2}(L_{\alpha}L_{\beta}+L_{\beta}L_{\alpha})-\delta_{\alpha \beta}\Vec{L}^2$ 
($\alpha$, $\beta$=$x$, $y$ and $z$), 
$\Vecsf{V}^\mathrm{out}$ the EFG tensor by the outside ions, 
and $V_{\alpha\beta}^\mathrm{out}$ 
the components of the EFG tensor, 
respectively. 

While $\mathcal{H}_\mathrm{mag}$ 
generates a single resonance peak for each of $^{47}$Ti and $^{49}$Ti, 
the EFG at Ti nuclei 
splits the spectrum to 2$I$ resonance peaks through $\mathcal{H}_\mathrm{el}$ 
in the case of $\mathcal{H}_\mathrm{el} \ll \mathcal{H}_\mathrm{mag}$. 
As seen in Fig. 1, the NMR spectra of $^{49}$Ti are split to seven peaks 
with almost the same frequency interval of $\sim$ 1.6 MHz, 
indicating $\mathcal{H}_\mathrm{el} \ll \mathcal{H}_\mathrm{mag}$. 
Since the EFG tensor $\Vecsf{V}^\mathrm{eff}$ 
is symmetric tensor, $\Vecsf{V}^\mathrm{eff}$ can be diagonalized 
by an appropriate rotation of the coordination system. 
Now we take the $x'y'z'$-coordination system to diagonalize 
$\Vecsf{V}^\mathrm{eff}$, and refer to the maximum value of 
the diagonal components as $V_{z' \! z'}^\mathrm{eff}$. 
The quadrupole frequency $\nu_Q$ 
is defined using $V_{z' \! z'}^\mathrm{eff}$ as 
\begin{eqnarray}
\nu_Q \equiv \frac{3eQV_{z' \! z'}^\mathrm{eff}}{2hI(2I-1)}. 
\end{eqnarray}
$\nu_Q$ increases 
with the deviation from cubic symmetry 
for EFG around the nuclei ($\nu_Q$=
$V_{z' \! z'}^\mathrm{eff}$=$V_{x' \! x'}^\mathrm{eff}$=$V_{y' \! y'}^\mathrm{eff}$=0 
in cubic symmetry 
because the trace of $\Vecsf{V}^\mathrm{eff}$ is 0). 
The splitting frequency $\Delta \nu$ is expressed 
by the first order perturbation theory of 
$\mathcal{H}_\mathrm{el}$, 
for example, in the case of the axially symmetric EFG, as
\begin{eqnarray}
\Delta \nu =\frac{3 \cos^2 \theta -1}{2} \nu_Q \le |\nu_Q|, 
\end{eqnarray}
where $\theta$ represents the angle 
between the z'-axis and ${\Vec H}^\mathrm{eff}$. 
The inequality $\Delta \nu \le |\nu_Q|$ is valid in general 
for $\mathcal{H}_\mathrm{el} \ll \mathcal{H}_\mathrm{mag}$. 
Therefore, since the present NMR spectra show 
$^{49}\Delta \nu$ is about 1.6 MHz, 
\begin{eqnarray}
|^{49}\nu_Q| \ge 1.6 \mbox{ MHz}. 
\end{eqnarray}
As mentioned above, $\nu_Q$ reflects 
the distortion of the EFG, i.e., the charge distribution. 
The EFG at Ti nuclei is contributed from 
the on-site 3$d$ electron and the outside 
O$^{2-}$, La$^{3+}$ and Ti$^{3+}$ ions. 
Generally, it is very difficult to estimate the contribution 
of outside ions to the EFG 
quantitatively, because
the EFG from the outside ions is largely enhanced by core electrons 
(the Sternheimer antishielding factor $\gamma_{\infty}$ corresponds 
to this effect). 
However, we could naively expect 
that the EFG originating from outside ions is small, 
since the distortion of TiO$_6$ octahedra is small in LaTiO$_3$. 

The above value of $|^{49}\nu_Q|$ is much larger 
than those of ATiO$_3$ compounds with Ti$^{4+}$(3$d^0$), 
in which the EFG is determined 
only by the contribution of the outside ions\cite{Padro,Bastow1}. 
It is reported that the value of $^{49}\nu_Q$ is 1.1 MHz for MgTiO$_3$, 
in which both of the crystalline and TiO$_6$ distortions are 
much larger than those in LaTiO$_3$\cite{Padro}. 
Padro \textit{et al.} systematically and quantitatively 
investigated the relationship between 
the EFG and the structural parameters for ATiO$_3$ compounds\cite{Padro}. 
They found that the shear strain of TiO$_6$ octahedra well correlates 
with the EFG magnitude. 
In accordance with their argument, 
we derived the shear strain $|\Psi|$ of LaTiO$_3$ 
as 0.27 using the structural data at 8 K in Ref.13. 
From this value of $|\Psi|$ 
we estimate the value of $^{49}\nu_Q$ 
contributed by the outside ions 
as $\sim$0.15 MHz at 8 K. 
Even if we consider the diffference of the valence of ions 
between La$^{3+}$Ti$^{3+}$O$_{3}$ and A$^{2+}$Ti$^{4+}$O$_{3}$, 
it is difficult to ascribe the origin of the large $\nu_Q$ value in LaTiO$_3$ 
to the outside ions. 
Therefore, the main contribution to the EFG at Ti nuclei 
should come from the on-site 3$d$ electron. 
So, we treat 
only $\mathcal{H}_\mathrm{el}^\mathrm{on}$ 
as the electric hyperfine interaction hereafter. 

The conclusion mentioned above means that the shape 
of 3$d$ orbital is deformed from cubic symmetry. 
The quadrupole frequency $\nu_Q$ is 
associated with the electron quadrupole moment $\Vecsf{q}$. 
We define $q_{z' \! z'}$ as the maximum value 
of the principal-axes components for $\Vecsf{q}$. 
We also introduce $r_\mathrm{el}$ defined as 
$r_\mathrm{el} \equiv \langle r^{-3} \rangle_\mathrm{el}
/\langle r^{-3} \rangle_\mathrm{FI}$, 
in which $\langle r^{-3} \rangle_\mathrm{FI}$ 
represents the $r^{-3}$ expectation value of the $3d$ electron 
calculated for a free Ti$^{3+}$ ion, 2.552 a.u.\cite{Abragam}. 
We could expect $r_\mathrm{el} < 1$ 
due to the hybridization with oxygen $2p$ states. 
Using Eqs. (4), (6), and (8), the following inequality is obtained 
for $q_{z' \! z'}$: 
\begin{eqnarray}
&\lefteqn{}& ^{49}\nu_Q = \frac{e^2Q}{7hI(2I-1)} 
r_\mathrm{el} \langle r^{-3} \rangle_\mathrm{FI} \,\, q_{z' \! z'}, \\
&\lefteqn{}& |q_{z' \! z'}| \equiv |3L^2_{z'} - {\Vec L}^2| \ge 1.62/r_\mathrm{el}. 
\end{eqnarray}

The value of $|q_{z' \! z'}|$ takes 6 
at the maximum for 3$d^1$ states, which is the case with 
the $d_{xy}$-type ($q_{z' \! z'}=6$) 
and $d_{3z^2-r^2}$-type ($q_{z' \! z'}=-6$) orbitals. 
The mixing among those orbitals reduces $|q_{z' \! z'}|$. 
Eq. (10) shows that the magnitude of 
the electron quadrupole moment $|q_{z' \! z'}|$ in LaTiO$_3$ is at least 
1.62/6 $\sim$ 27 \% of those for $d_{xy}$-type 
or $d_{3z^2-r^2}$-type orbitals. 
This result seems to be inconsistent with  
the proposed orbital liquid models\cite{KhaliullinMaekawa,Kikoin}, 
since it is supposed to give a cubicly symmetric charge distribution for 3$d$ electrons 
and very small $q_{z' \! z'}$. 
Thus, another orbital ordered state is expected to be realized in LaTiO$_3$. 

\begin{figure}[t]
\begin{center}\leavevmode
\includegraphics[width=0.77\linewidth]{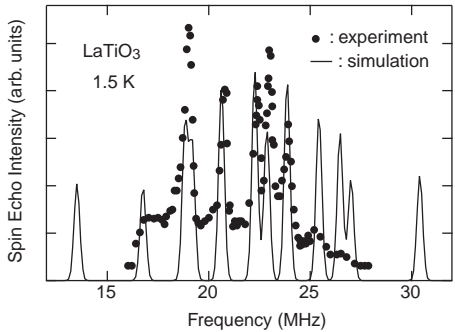}
\caption{Simulated and experimental $^{47,49}$Ti NMR spectra of LaTiO$_3$.
Solid circles represent experimental spectra, and 
solid lines represent the resonance spectra calculated for the 
$1/\sqrt{3}(d_{xy}+d_{yz}+d_{zx})$-like orbital ordering state. 
}\label{figure2}
\end{center}
\end{figure}

The neutron scattering measurements have shown that 
the ordered magnetic moment is 0.46 $\mu_B$ in the stoichiometric 
LaTiO$_3$\cite{Meijer}. 
Since the spin-wave approximation 
for a three dimensional $S$=1/2 model 
gives shrunken ordered spin moments of 0.84 $\mu_B$, 
the observed moment is about 0.4 $\mu_B$ smaller. 
The origin of this shrinkage is unclear. 
The present NMR spectra show that 
the internal magnetic field $|{\Vec H}^\mathrm{eff}|$ 
at Ti nuclei is 92 kOe. 
$|{\Vec H}^\mathrm{eff}|$ 
is expressed as 
\begin{eqnarray}
|{\Vec H}^\mathrm{eff}| &=& 
2 \mu_B r_\mathrm{mag} \langle r^{-3} \rangle_\mathrm{FI}
|\langle - \kappa {\Vec S} - \frac{2}{21} {\Vec S} \cdot \Vecsf{q} 
+ {\Vec L} \rangle| \nonumber \\
&=& 92 \mbox{ kOe}, \\
&\lefteqn{}&\!\!\!\!\!\!\!\!\!\!\!\!\!\!\!\!\!\!\!\!\!\!\!\! 
r_\mathrm{mag} |\langle - \kappa {\Vec S} - 
\frac{2}{21} {\Vec S} \cdot \Vecsf{q} 
+ {\Vec L} \rangle|
= 0.29, 
\end{eqnarray}
where $r_\mathrm{mag}$ represents $\langle r^{-3} \rangle_\mathrm{mag}
 / \langle r^{-3} \rangle_\mathrm{FI}$. 
The first term of Eq. (11) represents 
the Fermi contact field, the second the dipole field due to 
the electron spin moment, and the third the orbital field 
coming from the electron orbital moment. 
$r_\mathrm{mag} \times \kappa$  is supposed to be 
0.63 $\sim$ 0.75, 
since the isotropic hyperfine coupling constant $A_\mathrm{iso}$ 
(= $-159.7\times r_\mathrm{mag} \kappa$ [kOe/$\mu_B$] )
by the Fermi contact interaction is generally given as 
$-100$ $\sim$ $-120$ kOe/$\mu_B$\cite{MetallicShifts}. 
On the other hand, 
the ordered magnetic moment of 0.46 $\mu_B$ should correspond to 
\begin{eqnarray}
|  \langle 2{\Vec S} +{\Vec L} \rangle |=0.46.
\end{eqnarray}
So, Eqs. (12) and (13) must be satisfied at the same time. 
This restriction is so hard that it makes a good test 
for the models of the electronic states in LaTiO$_3$. 
As a rough estimation, if we assume that 
{\boldmath $S$} and {\boldmath $L$} are antiparallel, 
the orbital moment $|\Vec{L}|$ cannot exceed 0.29/$r_\mathrm{mag}$, 
since $|\kappa {\Vec S}|$ is larger than $2/21|{\Vec S}\cdot \Vecsf{q}|$ 
in the present $d^1$ case.

\begin{table}
\caption{\label{tab:table1}The values of the parameters $r_\mathrm{mag}$, 
$r_\mathrm{el}$ and $\kappa$ used 
for the NMR spectrum simulation based on the 
$1/\sqrt{3}(d_{xy}+d_{yz}+d_{zx})$-like orbital ordering model 
in LaTiO$_3$. 
The estimated values of 
$|\langle 2{\Vec S} \rangle|$ and $|\langle {\Vec L} \rangle|$ 
are also listed.} 
\begin{ruledtabular}
\begin{tabular}{ccccc}
$r_\mathrm{mag}$&
$r_\mathrm{el}$&$\kappa$&$|\langle 2{\Vec S} \rangle|$&$|\langle {\Vec L} \rangle|$\\
\hline
0.57&0.57&1.1&0.57&0.11\\%
\end{tabular}
\end{ruledtabular}
\end{table}

As noted above, the NMR spectra of LaTiO$_3$ show that 
a certain amount of the Ti 3$d$ electronic quadrupole moment 
is present. 
Mochizuki and Imada as well as Cwik \textit{et al.} proposed 
that 3$d$ electrons in LaTiO$_3$ occupy 
$1/\sqrt{3}(d_{xy}+d_{yz}+d_{zx})$-like orbitals 
and explained the physical properties of 
LaTiO$_3$ successfully\cite{Mochizuki2,Cwik}. 
This model is consistent with the present NMR results. 
So we simulated the NMR spectra assuming this type of orbital state. 
We assumed that the magnetic moment is parallel to the $a$-axis\cite{Cwik,Fritsch}. 
Furthermore, we set $A_\mathrm{iso}$ as $-100$ kOe/$\mu_B$ 
with the assumption that {\boldmath $S$} and {\boldmath $L$} are antiparallel. 
The resonance intensity 
was derived from the transition probability 
between the nuclear spin eigenstates.
The result is shown in Fig. 2. The simulated spectra well 
explain the experimental data. 
Especially, the resonance frequency is completely reproduced, 
while the deviation of the intensity is seen for outside peaks. 
The used parameters are listed in Table I. 
The 3$d$ orbital states of Ti(1) - Ti(4) sites in LaTiO$_3$ are represented as 
$ad_{xy}$$-$$bd_{yz}$$-$$cd_{zx}$, $ad_{xy}$$-$$cd_{yz}$$-$$bd_{zx}$, 
$ad_{xy}$+$bd_{yz}$+$cd_{zx}$ and $ad_{xy}$+$cd_{yz}$+$bd_{zx}$, 
respectively, where $a^2+b^2+c^2$=1. 
The four Ti sites give same spectra in the case of ${\Vec H}^\mathrm{eff}$ 
parallel to the $a$-axis. 
The orbital parameters are determined as 
$a$= 0.565, $b$= 0.452 and $c$= 0.690, 
which are close to the results in Refs.12 and 13. 
Figure 3 shows the orbital states in LaTiO$_3$ expected from these parameters. 
It is important that $\nu_Q$ 
is led 
from the electron quadrupole moment with the reasonable parameters. 
The present wave functions and parameters give $q_\mathrm{z' \! z'}$= $-5.77$, 
$^{49}\nu_Q=$ 3.2 MHz, and $\eta = 0.49$. 
The reduction of $r_\mathrm{mag}$ and $r_\mathrm{el}$ 
from 1 partly originates from the hybridization 
between Ti 3$d$ and O 2$p$ orbitals, 
although the values of 0.57 for $r_\mathrm{mag}$ and $r_\mathrm{el}$ appear rather small. 
It might be affected by the shielding effect of core electrons 
which can modulate the EFG from on-site $d$ electrons 
by 10$-$20 \%\cite{Abragam}. 
Possibly orbital fluctuation might also contribute the reduction. 
The orbital magnetic moment $\mu_B \Vec{L}$ is estimated as 0.11 $\mu_B$, 
which may be overestimated 
by the underestimation of $r_\mathrm{mag}$. 
If we vary $A_\mathrm{iso}$ between $-80$ and $-120$ kOe/$\mu_B$, 
estimated $\mu_B \Vec{L}$ ranges from $0.05$ to $0.15$ $\mu_B$. 
In order to be more accurate at 
the estimation of $r_\mathrm{mag}$ and $r_\mathrm{el}$, 
more detailed analyses 
such as TiO$_6$ cluster model calculations 
will be needed. 

\begin{figure}[t]
\begin{center}\leavevmode
\includegraphics[width=0.77\linewidth]{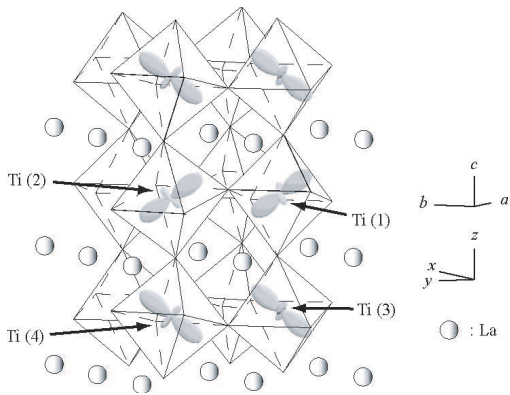}
\caption{ Orbital states expected by 
the $1/\sqrt{3} (d_{xy}+d_{yz}+d_{zx})$-like orbital ordering model 
in LaTiO$_3$. 
The detailed representation of the wave function for Ti 3$d$ electrons 
is described in the text. 
}\label{figure3}
\end{center}
\end{figure}

Some other approaches are taken to explain 
the magnetism of LaTiO$_3$ 
such as a strong spin-orbit interaction\cite{Hemberger,Fritsch}. 
The degeneracy of $\tilde{j}$= 3/2 quadruplet 
is lifted to two doublets 
by trigonal crystalline distortion. 
The resultant ground states show magnetic moment 
through hybridization with $\tilde{j}$= 1/2. 
Now we consider in the ground doublet a state 
which reproduces the observed magnetic moment of 0.46 $\mu_B$ 
parallel to the $a$-axis. 
This state exhibits an anisotropic orbital shape elongated 
in the (1,1,1)-direction with $q_\mathrm{z' \! z'}$ $\sim$$-5.0$, 
which is comparable with the orbital ordering model\cite{Mochizuki2,Cwik}, 
while it gives rather large orbital moment $\sim$0.44 $\mu_B$. 
Based on this type of orbital state, 
the orbital magnetic moment would be expected to take 
0.05 $\sim$ 0.2 $\mu_B$ from Eqs. (12) and (13) 
in the conceivable range of hyperfine coupling parameters 
$r_\mathrm{mag}$ and $\kappa$. 
This fact seems to suggest that 
the effect of the trigonal crystalline distortion is rather large 
compared with the SO interaction 
as is pointed out by Cwik \textit{et al.}\cite{Cwik}, and 
that some part of the reduced magnetic moment 
should be ascribed to other reasons. 
Mochizuki and Imada have accounted for the reduction 
by the itinerancy of the $t_{2g}$ electrons, 
i.e., the double occupation of up and down spin electrons 
in the $t_{2g}$ state\cite{Mochizuki2}. 
It seems to support this account that the increase of 
oxygen content gradually decreases the ordered moment 
and finally leads to a Pauli paramagnetic state. 

In summary, we analyzed the $^{47,49}$Ti NMR spectra of LaTiO$_3$. 
The NMR technique can evaluate the electron quadrupole moment $\Vecsf{q}$, 
i.e., the order parameter of orbital ordering, quantitatively. 
The NMR spectra show that the electron quadrupole moment is rather large. 
It is supposed that LaTiO$_3$ takes an orbital ordering state 
represented approximately as  $1/\sqrt{3} (d_{xy}+d_{yz}+d_{zx})$. 

This work was supported by the Yamada Science Foundation and 
a Grant-in-Aid for Scientific Research 
from the Ministry of Education, Culture, Sports, Science and Technology of Japan.

\bibliography{text}

\begin{thebibliography}{9}
\bibitem{Tokura-Nagaosa} Y. Tokura and N. Nagaosa, Science {\bf 288}, 462 (2000).
\bibitem{Keimer} B. Keimer \textit{et al.}, Phys. Rev. Lett. {\bf 85}, 3946 (2000).
\bibitem{KhaliullinMaekawa} G. Khaliullin and S. Maekawa, Phys. Rev. Lett. {\bf 85}, 3950 (2000).
\bibitem{Kikoin} K. Kikoin, O. Entin-Wohlman, V. Fleurov, and A. Aharony, Phys. Rev. {\bf B 67}, 214418 (2003).
\bibitem{KhaliullinOkamoto} G. Khaliullin and S. Okamoto, Phys. Rev. Lett. {\bf 89}, 167201 (2002). 
\bibitem{Ulrich} C. Ulrich \textit{et al.}, Phys. Rev. Lett. {\bf 89}, 167202 (2002). 
\bibitem{Mizokawa} T. Mizokawa and A. Fujimori, Phys. Rev. {\bf B 54}, 5368 (1996).
\bibitem{Sawada} H. Sawada and K. Terakura, Phys. Rev. {\bf B 58}, 6831 (1998).
\bibitem{Akimitsu} J. Akimitsu \textit{et al.}, J. Phys. Soc. Jpn. {\bf 70}, 3475 (2001).
\bibitem{Itoh1} M. Itoh, M. Tsuchiya, H. Tanaka, and K. Motoya, 
J. Phys. Soc. Jpn. {\bf 68}, 2783 (1999); M. Itoh and M. Tsuchiya, 
J. Magn. Magn. Mater. {\bf 226-230}, 874 (2001).
\bibitem{Nakao} H. Nakao \textit{et al.}, Phys. Rev. {\bf B 66}, 184419 (2002).
\bibitem{Mochizuki2} M. Mochizuki and M Imada, J. Phys. Soc. Jpn. {\bf 70}, 2872 (2001); M. Mochizuki and M Imada, cond-mat/0301049.
\bibitem{Cwik} M. Cwik \textit{et al.}, Phys. Rev. {\bf B 68}, 060401 (2003).
\bibitem{Hemberger} J. Hemberger \textit{et al.}, Phys. Rev. Lett. {\bf 91}, 066403 (2003). 
\bibitem{Meijer} G. I. Meijer \textit{et al.}, Phys. Rev. {\bf B 59}, 11832 (1999). 
\bibitem{MetallicShifts} \textit{Metallic Shifts in NMR}, edited by G. C. Cater, 
L. H. Bennet, and D. K. Kahan (Pergamon, New York, 1977).
\bibitem{Furukawa} Y. Furukawa \textit{et al.}, Phys. Rev. {\bf B 59}, 10550 (1999). 
\bibitem{Abragam} A. Abragam and F. Bleany, \textit{Electron Paramagnetic Resonance of Transition Ions} (Clarendon Press, Oxford, 1970). 
\bibitem{Padro} D. Padro, A. P. Howes, M. E. Smith, and R. Dupree, Solid State NMR {\bf 15}, 231 (2000). 
\bibitem{Bastow1} T. J. Bastow, M. A. Gibson, and C. T. Forwood, Solid State NMR {\bf 12}, 201 (1998); T. J. Bastow and H. J. Whitfield, Solid State Commun. {\bf 117}, 483 (2001). 
\bibitem{Fritsch} V. Fritsch \textit{et al.}, Phys. Rev. {\bf B 65}, 212405 (2002). 
\end{thebibliography}

\end{document}